\documentclass[twocolumn,english,superscriptaddress]{revtex4-1}
\usepackage[T1]{fontenc}
\usepackage{amsfonts}
\usepackage{amsmath}
\usepackage[latin9]{inputenc}
\usepackage{textcomp}
\usepackage{amstext}
\usepackage{subscript}
\usepackage{graphics}
\usepackage{graphicx}
\usepackage{float}

\makeatletter

\newcommand{\lyxmathsym}[1]{\ifmmode\begingroup\def\b@ld{bold}
  \text{\ifx\math@version\b@ld\bfseries\fi#1}\endgroup\else#1\fi}

\usepackage{hyperref}
\usepackage{upgreek,textgreek}
\hypersetup{colorlinks=true,linkcolor=blue,citecolor=blue,urlcolor=blue}
\renewcommand{\fnum@figure}{Fig.~\thefigure}
\makeatother

\usepackage{babel}
\begin{document}
\title{Relevance of sample geometry on the in-plane anisotropy of Sr\textsubscript{$x$}Bi\textsubscript{2}Se\textsubscript{3}
superconductor}
\author{Xinrun Mi}
\affiliation{Low Temperature Physics Lab, College of Physics \& Center of Quantum
Materials and Devices, Chongqing University, Chongqing 401331, China}
\author{Yecheng Jing}
\affiliation{Low Temperature Physics Lab, College of Physics \& Center of Quantum
Materials and Devices, Chongqing University, Chongqing 401331, China}
\author{Kunya Yang}
\affiliation{Low Temperature Physics Lab, College of Physics \& Center of Quantum
Materials and Devices, Chongqing University, Chongqing 401331, China}
\author{Yuhan Gan}
\affiliation{Low Temperature Physics Lab, College of Physics \& Center of Quantum
Materials and Devices, Chongqing University, Chongqing 401331, China}
\author{Aifeng Wang}
\affiliation{Low Temperature Physics Lab, College of Physics \& Center of Quantum
Materials and Devices, Chongqing University, Chongqing 401331, China}
\author{Yisheng Chai}
\affiliation{Low Temperature Physics Lab, College of Physics \& Center of Quantum
Materials and Devices, Chongqing University, Chongqing 401331, China}
\author{Mingquan He}
\email{mingquan.he@cqu.edu.cn}

\affiliation{Low Temperature Physics Lab, College of Physics \& Center of Quantum
Materials and Devices, Chongqing University, Chongqing 401331, China}
\date{today}

\begin{abstract}
Possible emergence of nematic superconductivity that breaks its underlying
lattice symmetry in doped topological insulator Bi\textsubscript{2}Se\textsubscript{3,}establishes
this system as a unique candidate of topological superconductors. Exclusion
of possible extrinsic causes for the two-fold superconductivity is
essential to clarify its topological nature. Here, we present electrical
transport on Sr\textsubscript{$x$}Bi\textsubscript{2}Se\textsubscript{3}
superconductors with rectangular and circular geometries. The occurrence of the two-fold symmetric in-plane
upper critical field $H_{c2}$ is found to be weakly geometry dependent.
However, the anisotropic ratio between the maximum and minimum in-plane
upper critical fields varies significantly among samples with different
shapes. Compared with the rectangular sample, the anisotropic ratio is largely suppressed in the circular sample which has higher geometric rotational symmetry.  Our results imply that sample geometry plays a subdominant role, but circular shape is more ideal to reveal the two-fold superconductivity of Sr$_x$Bi$_2$Se$_3$ in the vortex state.
\end{abstract}

\maketitle

\section{Introduction}

The emergence of spontaneous symmetry breaking in addition to $U(1)$
gauge symmetry in superconductors often leads to unconventional superconductivity,
such as $d$-wave, chiral $p$-wave superconductivity. One more prominent
example can be found in a nematic superconductor with odd-parity pairing
symmetry, which breaks the lattice rotational symmetry in the superconducting
state \cite{Fu:2010aa,Fu:2014aa}. Even more intriguing of such a nematic superconductor is its topological nature, which could realize the time-reversal-invariant topological superconducting state \cite{Fu:2010aa,Fu:2014aa}. Expectations of finding Majorana fermion excitation in the nematic topological superconductors have attracted much research efforts in the community of condense matter physics \cite{Yonezawa:2019aa}.   

Material realization of the nematic superconductor is made possible by the discoveries of superconductivity in Cu, Sr or Nb doped topological insulator Bi\textsubscript{2}Se\textsubscript{3}. Soon after the report on superconductivity of Cu intercalated Bi\textsubscript{2}Se\textsubscript{3} \cite{Hor:2010}, F. Liang \textit{et al.} \cite{Fu:2010aa} suggested the possibility of realizing an odd-parity superconducting state in this system. Inspired by the observation of an unusual two-fold symmetric spin susceptibility that breaks the trigonal lattice symmetry of Cu$_x$Bi\textsubscript{2}Se\textsubscript{3} in the superconducting state \cite{Matano2016}, the 'nematic superconductor' hosting a two-dimensional $E_u$ odd-pairing state was proposed by F. Liang to explain the unconventional symmetry breaking effects \cite{Fu:2014aa}.  Since then, numerous studies have reported signatures of nematic superconductivity in $A_x$Bi\textsubscript{2}Se\textsubscript{3} (A= Cu, Sr or Nb) using various techniques including electrical transport \cite{Nikitin2016,Pan2016,Du2017,Shen2017,Smylie2018}, specific heat \cite{Yonezawa2017,Willa2018,Sun2019}, magnetic torque \cite{Tomoya_2017}, thermal-expansion \cite{Cho2020}, scanning tunneling microscope (STM) \cite{Ran_2018}. However, to which extent the observed two-fold symmetric superconductivity is originated from an intrinsic topological nature still remains elusive. Other factors, such as built-in strain or distortion \cite{Kuntsevich_2018,Kuntsevich_2019}, structural inhomogeneity and impurity phases \cite{Shruti2015,Kaya_2017,Kamminga2020,Simone_2021}, inhomogeneously distributed superconducting regions \cite{Hor:2010,Kriener2011,Levy_2013,Mann2014,Ran_2018}, cannot be completely ruled out at this stage. Exclusion of possible extrinsic origins is of great importance to clarify the putative topological nature of the nematic superconductivity found in doped Bi\textsubscript{2}Se\textsubscript{3} systems.       

Many experimental identifications of nematic superconductivity in $A_x$Bi\textsubscript{2}Se\textsubscript{3} were performed in the vortex state by applying magnetic fields that rotate in-plane \cite{Yonezawa:2019aa}. Various factors could affect the in-plane anisotropy in the vortex state especially for crystals with irregular shapes. Firstly, the demagnetization effect exists inevitably in any samples with finite sizes, which could vary substantially along different in-plane magnetic field directions except for circular or spherical shapes. The geometric effects can have further impact on the flux penetration, in a way that the penetration of flux lines at sharp corners is more difficult compared to smooth areas  \cite{Brandt2001}. One more possibility could be the influence of surface superconductivity, considering the weakly van-der Waals coupled layer structure of $A_x$Bi\textsubscript{2}Se\textsubscript{3}. As already pointed out by Saint-James and de Gennes in 1963 \cite{SAINTJAMES1963306}, for the in-plane field configuration,  nucleation of superconductivity on the surfaces enables a higher surface critical field $H_{c3}=1.69H_{c2}$ than that of the bulk $H_{c2}$. Importantly, the $H_{c3}$ depends on sample geometry and substantial enhancement of $H_{c3}$ to values higher than $1.69H_{c2}$ is expected at edges or vertices \cite{Fomin1998,Lu2000,Pan2002}. Thus, different in-plane magnetic field configurations with respect to irregular shapes containing various sharp edges may significantly affect the in-plane anisotropic superconducting properties. Detail analysis of the impacts induced by sample geometry on the nematic superconductivity is, however, rarely explored.      
\begin{figure*}[t]
\centering
\includegraphics[scale=0.7]{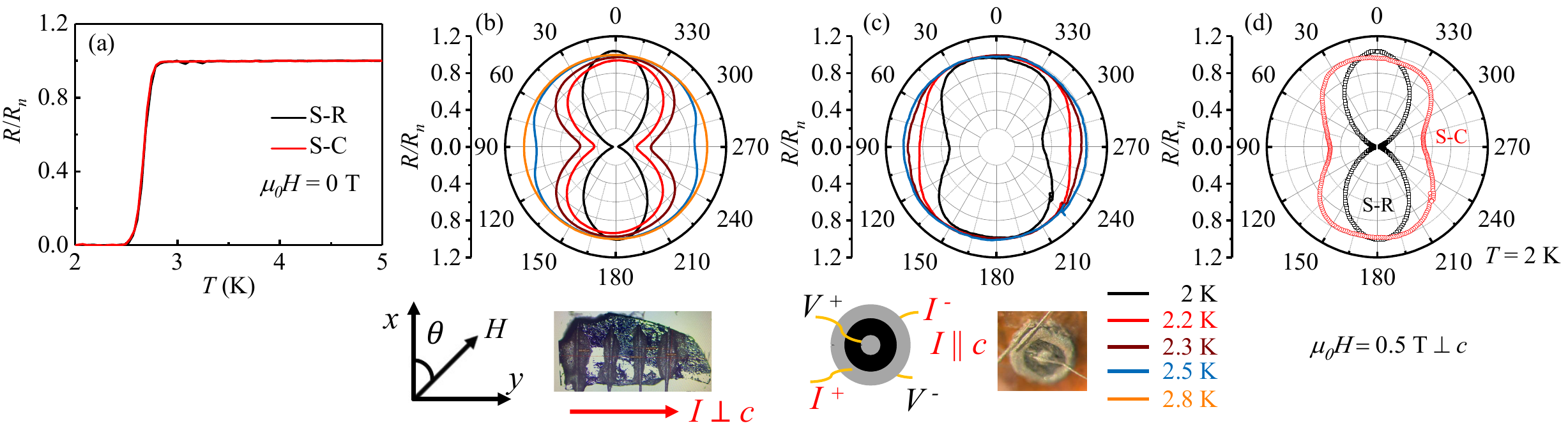}
\caption{(a) Normalized electric resistance $R/R_n$(5 K) of the rectangular (S-R) and circular samples (S-C). (b)(c) The angular variation of resistance $R(\theta)$ of S-R and S-C samples measured near $T_c$ in a fixed in-plane magnetic field $\mu_0H$=0.5 T. The photographs in (b) and (c) are optical images of S-R and S-C. The current is along the in-plane direction for S-R, which is perpendicular to the magnetic field $I\perp H$ at $\theta=0^\circ$.  The current was applied out-of-plane and the Corbino-like electrodes were used for S-C.   (d) Comparison of $R(\theta)$ recorded in $\mu_0H$=0.5 T for S-R and S-C at 2 K.} 
\label{fig1}
\end{figure*}

In this article, we demonstrate the impacts of sample geometry  on the in-plane anisotropic superconductivity of  Sr$_x$Bi$_2$Se$_3$. We managed to engineer one single piece of crystal into different shapes including rectangular bar and circular disk, which enables direct investigation of geometric effects.  It is found that the in-plane anisotropic magneto-transport properties in the vortex state depend strongly on crystal shapes, but the two-fold symmetry remains generic.  The anisotropic ratio between the maximum and minimum in-plane $H_{c2}$ varies with sample shapes, which appears to be much smaller in the circular sample compared with that in  rectangular sample. Our findings suggest that sample geometry does affect the in-plane anisotropy of Sr$_x$Bi\textsubscript{2}Se\textsubscript{3}, and that it is better to use circular geometry when studying anisotropic superconducting properties in the vortex state.

\section{Methods}
 
Single crystals of Sr$_x$Bi\textsubscript{2}Se\textsubscript{3} with nominal strontium $x$=0.2 were grown by the melt-growth method \cite{Liu2015}. Stoichiometric mixture of high purity Sr piece$(99.5\%)$, Bi powder$(99.999\%)$ and Se powder$(99.999\%)$ was sealed in a evacuated quartz tube. Then, the quartz tube was heated to 850 $^\circ$C and kept at this temperature for 48 h. After that, the quartz tube was cooled to 610 $^\circ$C slowly with 3 $^\circ$C /h followed by fast quenching in water.
The in-plane magnetic field angle dependent electrical transport measurements were performed in a Physical Property Measurement System (PPMS, Quantum Design Dynacool 9 T), using a home-made insert equipped with a rotator. The electric resistance of samples with rectangular and circular shapes was measured using standard four-probe and Corbino-like methods, respectively. 

\section{Results And Discussions}

Figure \ref{fig1}(a) presents the zero field electric resistance of a rectangular bar shaped crystal ( S-R) and a circular disk sample (S-C), in the vicinity of the superconducting transition. The resistance has been normalized to its normal state value $R_n$($T$=5 K) for better comparison.  Both samples belong to a single piece of crystal. The circular sample was cut from the rectangular sample, and sharp edges were polished away subsequently.  The current $I$ was applied in-plane (out-of-plane) for the rectangular (circular) sample [see Figs. \ref{fig1}(b)(c)].  Nearly identical transitions are found in both samples, which show zero resistance below $T_c^{zero}$=2.5 K. The sample superconducting $T_c$ allows us to compare the geometrical effects directly in the coming discussions.    

\begin{figure*}[t]
\centering
\includegraphics[scale=0.6]{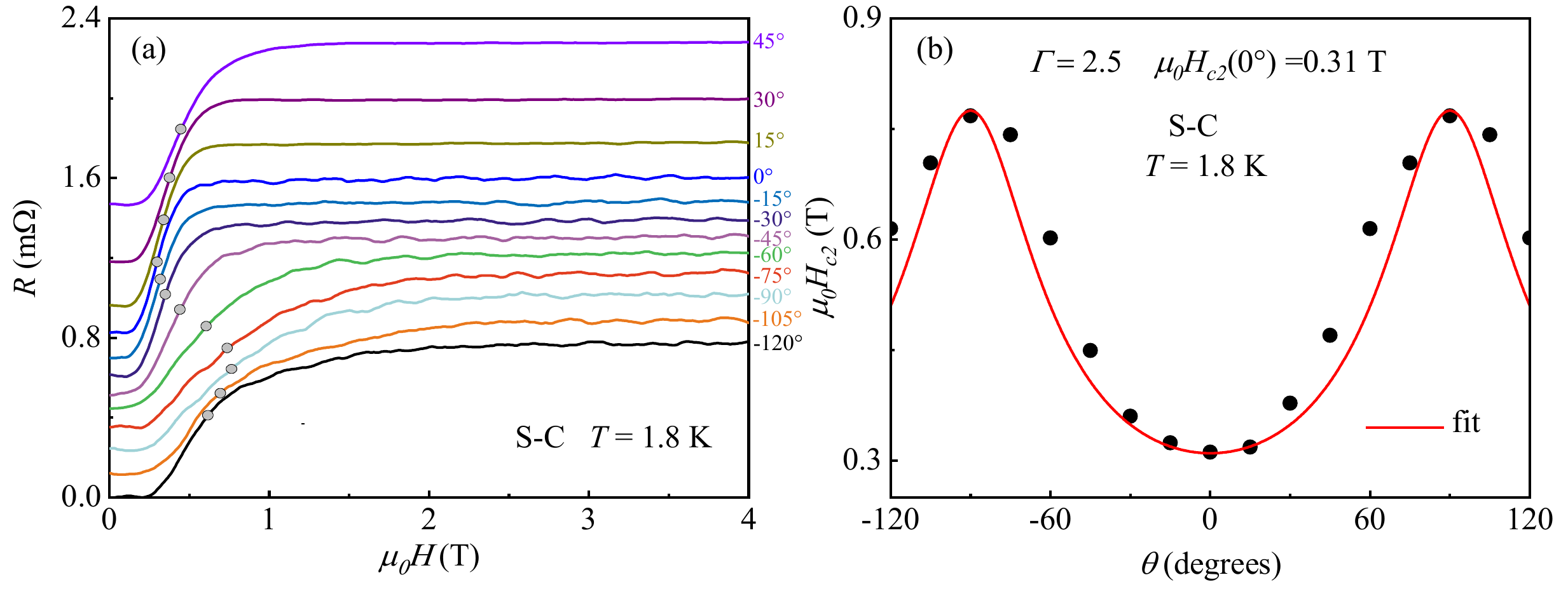}
\caption{(a) Magnetoresistance of the S-C sample measured at 1.8 K with magnetic field applied along various in-plane directions. The upper critical fields $H_{c2}$ are identified as the position where the magnetoresistance reaches 50\% of the normal state resistance, as marked out by gray dots.  (b) Angular variation of the upper critical field $H_{c2}(\theta)$. The red solid line is a theoretical description according to Eq. \ref{eqHc2} for an anisotropic superconductor. }
\label{fig2}
\end{figure*}

In Figs. \ref{fig1}(b)(c), we present the angular dependence of the resistance $R(\theta)$, recorded in a fixed field $\mu_0 H$=0.5 T with magnetic field rotating within the basal plane ($H\perp c$), for the rectangular and circular samples.    Clear two-fold symmetry is observed in $R(\theta)$ below $T_c$  for all samples, irrespective of the different geometries, suggesting a subdominant role played by the crystal shape. Despite the generic two-fold symmetry, the geometric effects still leave some fingerprints on $R(\theta)$. One can see that the anisotropy of $R(\theta)$ is more evident in the rectangular sample than that in the circular sample. This difference is seen more clearly in Fig. \ref{fig1}(d), which directly compares $R(\theta)$ of the two samples at 2 K. The anisotropic ratio between the maximum and minimum resistance is $\alpha =R_{max}/R_{min}$= 35.7 for the rectangular sample, which becomes much smaller in the circular ($\alpha=2$) sample. In addition, the two-fold symmetry disappears at 2.5 K in the circular sample, which is still visible in the rectangular sample at the same temperature. These results imply that different geometries do have impacts on the in-plane anisotropy, although the two-fold symmetry remains generic. We will come back to the discussion of the geometric effects in more detail later.    

The emergence of two-fold anisotropic $R(\theta)$ in the superconducting state points to a two-fold symmetric upper critical field $H_{c2}(\theta)$. Fig. \ref{fig2}(a) shows the magnetoresistance of the circular sample measured at 2 K with magnetic field pointing to various fixed angles. Clear anisotropy is seen, in which the superconductivity is suppressed in lower (higher) magnetic fields at $\theta=0^{\circ}$ ( $\theta=90^{\circ}$),  compared with other angles.  The angular variation of $H_{c2}(\theta)$ can be better visualized in in Fig. \ref{fig2}(b). The values of $H_{c2}(\theta)$ are determined according to the criteria when the magnetoresistance reaches 50\% of the normal state resistance [see the dots in Fig. \ref{fig2}(a)].  The two-fold symmetry of  $H_{c2}(\theta)$ is obvious, which suggests a similar profile of the superconducting gap $\Delta(\theta)$, considering the direct link between  $H_{c2}$ and $\Delta$. Indeed, the angular variation of $H_{c2}(\theta)$ can be traced nicely by [the solid line in Fig. \ref{fig2}(b)]:
\begin{equation}
H_{c2}(\theta)=\frac{H_{c2}(0^{\circ})}{\left(\cos^{2}\theta+\varGamma^{-2}\sin^{2}\theta\right)^{1/2}},
\label{eqHc2}
\end{equation} suggesting the anisotropic nature of the superconducting order parameter \cite{Pan2016,Shen2017}. Here, the anisotropic ratio $\varGamma= H_{c2}(90^{\circ})/H_{c2}(0^{\circ})$=2.5 and $\mu_0H_{c2}(0^{\circ})$=0.31 T give the best fit at 1.8 K.  Signature of three-fold symmetry of the underlying lattice is invisible, which might be the consequence of nematic superconductivity although the exact origin is still elusive \cite{Yonezawa:2019aa}. 

\begin{figure*}[t]
\centering
\includegraphics[scale=0.8]{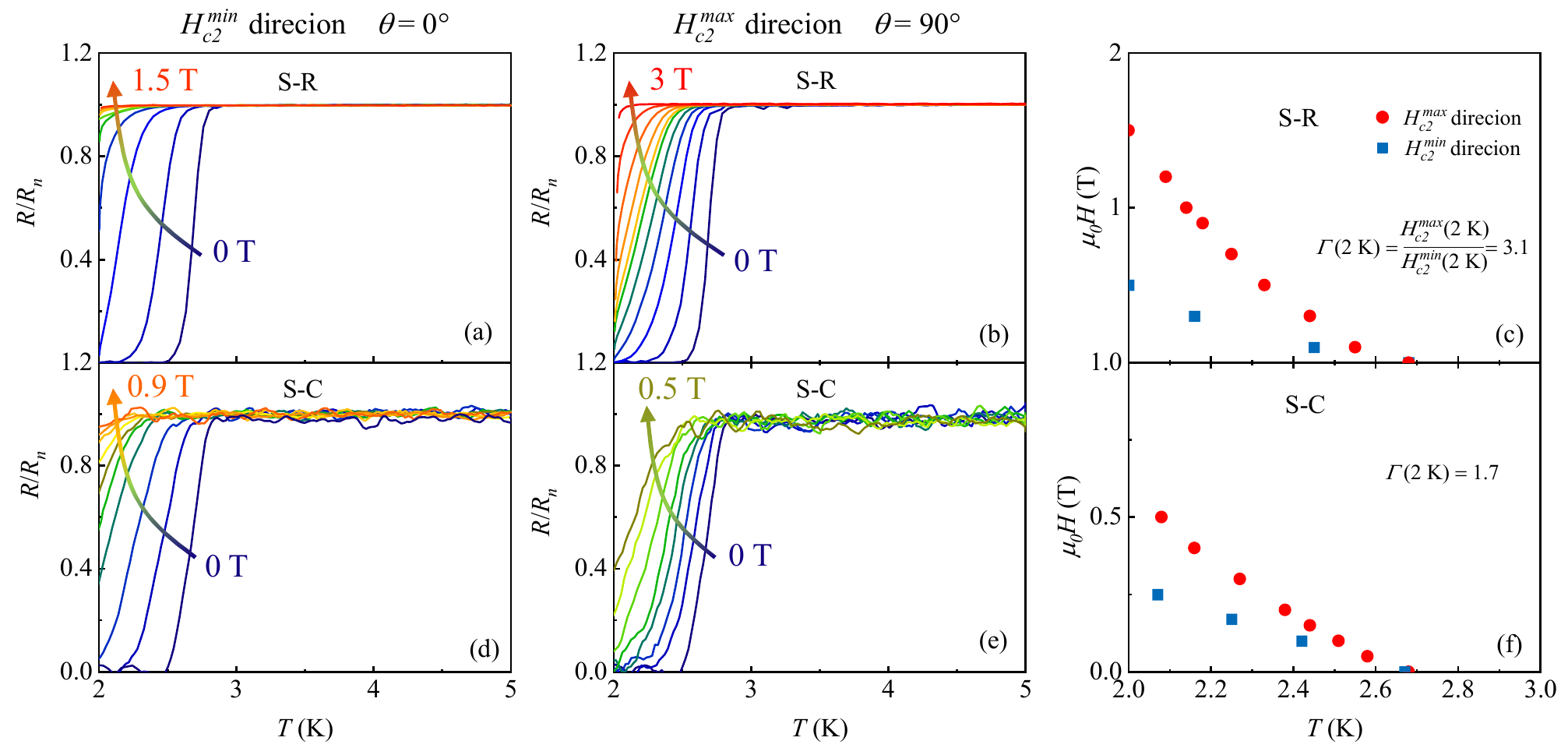}
\caption{ Temperature dependence of the resistance measured at fixed fields applying along (a)(d) the  minimum  [$H_{c2}^{min}(\theta=0^{\circ}$)], and (b)(e) the maximum [$H_{c2}^{max}(\theta=90^{\circ}$)]  upper critical fields directions for S-R (a)(b) and S-C (d)(e) samples. (c)(f) The anisotropic upper critical field determined from the data shown in (a)(b) and (d)(e).}
\label{fig3}
\end{figure*}

In Figure \ref{fig3}, we explore the anisotropic in-plane upper critical field in more detail. Now the magnetic field is directing along fixed angles where $H_{c2}$ reaches minimum [$H_{c2}^{min}(\theta=0^{\circ}$), Figs. \ref{fig3}(a)(d)] and maximum [$H_{c2}^{max}(\theta=90^{\circ}$), Figs. \ref{fig3}(b)(e)] values. From these results, the $H-T$ phase diagrams can be obtained, as shown in Figs. \ref{fig3} (c)(f). Again, the criteria where the resistance reaches 50\% of the normal state value is chosen for $H_{c2}(T)$. Clearly, significant anisotropy is found and the superconductivity is suppressed faster along the $H_{c2}^{min}(T)$ line than that of the $H_{c2}^{max}(T)$ boundary in both samples. The separation between $H_{c2}^{min}(T)$ and $H_{c2}^{max}(T)$ becomes largely reduced, when switching from the rectangular sample to the circular sample. At 2 K, the anisotropic ratio of the in-plane upper critical field, $\varGamma=H_{c2}^{max}/H_{c2}^{min}$, reads $\varGamma=$ 3.1 and 1.7 for the rectangular and circular samples, respectively.  Compared with the rectangular sample, $\varGamma$ is reduced by 45$\%$ in the circular sample, suggesting sizable contributions from the shape anisotropy.  

\begin{figure*}[t]
\centering
\includegraphics[scale=0.7]{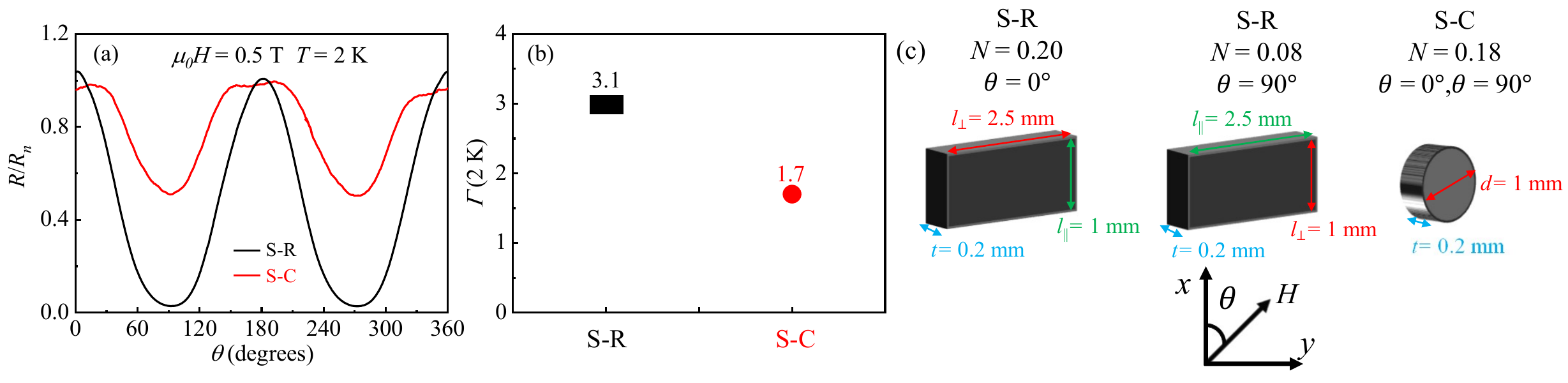}
\caption{Comparison of the (a) angular dependence of resistance $R_\theta$ and (b) anisotropic ratio $\varGamma$(2 K) of the upper critical field between the S-R and S-C samples. (c) Estimation of the demagnetization factors $N$ of the S-R and S-C samples when the magnetic field is applied along $\theta=0^\circ$ and $\theta=90^\circ$ directions. For S-R, the length parallel (perpendicular) to the field direction is defined as $l_\parallel$ ($l_\perp$). The diameter of the S-C sample is $d$. The thicknesses of both samples are labeled as $t$. }
\label{fig4}
\end{figure*}

In Figure \ref{fig4}, we compare the geometric effects directly using the data presented in Fig. \ref{fig1} and Fig. \ref{fig3}. As shown in Figs. \ref{fig4}(a), the angular variation of $R(\theta)$ shows strong sample dependence when other conditions are kept the same ($T$=2 K, $\mu_0H$=0.5 T). Only a single periodicity of 180$^\circ$ is found in both samples with the same maximum and minimum positions,  indicating negligible contributions from multi-domain structures. This suggests that the observed variations among different samples are mainly coming from geometry effects. The anisotropy is suppressed significantly in the circular sample, as also seen in the anisotropic ratio $ \varGamma$ displayed in Fig. \ref{fig4}(b). The anisotropic superconducting properties are likely amplified by the shape anisotropy in the rectangular sample.

For a finite sized sample with an ellipsoidal shape, the internal field $H_{in}$ experienced by the sample differs from the external applied field $H_0$ by:
\begin{equation}
H_{in}=H_0-NM,
\label{eq1}
\end{equation}where $N$ is the demagnetization factor, $M$ is the magnetization. Superconductors expel magnetic flux, and $M<$0 in the superconducting state, which gives a larger internal field than the field applied $H_{in}>H_0$. The demagnetization effects are rather complicated for other shapes. Following 
R. Prozorov \textit{et al.} \cite{Prozorov}, the effective demagnetization factors for a rectangular cuboid  and a circular disk can be approximated as:
\begin{equation}
N^{-1}=1+\frac{3}{4}\left(\frac{l_{\parallel}}{l_\perp}+\frac{l_\parallel}{t}\right),
\label{eq2}
\end{equation}
and
\begin{equation}
N^{-1}=2+\frac{1}{\sqrt{2}}\frac{d}{t},
\label{eq3}
\end{equation}
respectively. For the rectangular cuboid, the symbol $l_\parallel $ is the length along the field direction, $l_\perp $ is the size perpendicular to the magnetic field. The parameter $d$ is the diameter of the circular disk, and $t$ represents the thickness in both shapes [see also Fig. \ref{fig4}(c)]. Rough estimations of the demagnetization factors of our samples in certain field directions are shown in Fig. \ref{fig4}(c). Simplifications have been made here assuming ideal rectangular and circular geometries. One can see that, for the rectangular shape, the demagnetization factors differ by a factor of 2.5 between $\theta=0^\circ$ and $\theta=90^\circ$ directions. Such an geometrical anisotropy is absent in the circular shape at any in-plane directions. Along $\theta=0^\circ$, the demagnetization factor $N$ is similar in both shapes. For $\theta=90^\circ$, on the other hand, $N$ is much smaller in the rectangular shape than that in the circular geometry. Consequently, the rectangular sample experiences a smaller internal magnetic field $H_{in} (\theta=90^\circ)$, and superconductivity can survive up to a larger upper critical field identified by the externally applied field. The anisotropic ratio in the rectangular sample ($\varGamma_{S-R}/2.5=1.24$) actually gets close to that in the circular sample ($\varGamma_{S-C}=1.7$), if the anisotropic demagnetization effects are taken into account despite our oversimplified model. Note that the effect of demagnetization is overestimated here since $-M<H_0$ in the vortex state.  In addition to the anisotropic demagnetization factor, the sharp edges or vertices in the rectangular shape can delay the entrance of vortices \cite{Brandt2001}, or even accumulate higher surface upper critical fields \cite{Fomin1998,Lu2000,Pan2002}. These effects are hard to quantify, but certainly complicates the anisotropy further. One more possibility is the influence of Lorenz force, which is absent only for $H\parallel I$ at $\theta=90^\circ$ in the rectangular sample. The anisotropy brought by Lorenz force is minimized in the circular sample with current flowing along the $c$-axis. All these factors contribute to the observed different anisotropic ratios in different geometries. The circular shape appears to be more suitable to probe the intrinsic anisotropy.

\section{Conclusions}
In summary, we have investigated the geometric effects on the in-plane anisotropy of the nematic superconductor Sr$_x$Bi$_2$Se$_3$. The two-fold symmetry of the superconducting state appears to be generic, irrespective to samples with different shapes. The anisotropy, on the other hand, varies significantly with sample geometries. Compared with the rectangular sample, the anisotropic ratio of the in-plane upper critical field is largely reduced in the circular sample. Both anisotropic demagnetization factors, flux and/or surface superconductivity accumulation at sharp vertices,  may affect the in-plane anisotropy in samples with anisotropic shapes. Our findings suggest that the sample geometry plays a subdominant role in the anisotropic superconductivity, but can not be neglected completely in the study of nematic superconductivity in the vortex state.

\section*{Acknowledgments}

We thank Guiwen Wang and Yan Liu at the Analytical and Testing Center of
Chongqing University for technical support. This work has been supported
by National Natural Science Foundation of China (Grant Nos.11904040,
12047564), Chongqing Research Program of Basic Research and Frontier
Technology, China (Grant No. cstc2020jcyj-msxmX0263), Fundamental
Research Funds for the Central Universities, China(2020CDJQY-A056,
2020CDJ-LHZZ-010, 2020CDJQY-Z006), Projects of President Foundation
of Chongqing University, China(2019CDXZWL002). Y. Chai acknowledges
the support by National Natural Science Foundation of China (Grant
No. 11674384, 11974065). A. Wang acknowledges the support by National
Natural Science Foundation of China (Grant No. 12004056).

\bibliographystyle{apsrev4-1}

\end{document}